\documentclass[twocolumn, pre]{revtex4}

\usepackage{amssymb}
\usepackage{amsmath}
\usepackage{graphics}
\usepackage{graphicx}
\usepackage{mathrsfs}

\newcommand{\be}{\begin{equation}}
\newcommand{\ee}{\end{equation}}

\begin{document}

\title{Capacitance of the Double Layer Formed at the Metal/Ionic-Conductor Interface: How Large Can It Be?}

\date{\today}
\author{Brian Skinner}
\author{M. S. Loth}
\author{B. I. Shklovskii}
\affiliation{Fine Theoretical Physics Institute, University of Minnesota, Minneapolis, Minnesota 55455}

\begin{abstract}

The capacitance of the double layer formed at a metal/ionic-conductor interface can be remarkably large, so that the apparent width of the double layer is as small as 0.3 \AA.  Mean-field theories fail to explain such large capacitance. We propose an alternate theory of the ionic double layer which allows for the binding of discrete ions to their image charges in the metal. We show that at small voltages the capacitance of the double layer is limited only by the weak dipole-dipole repulsion between bound ions, and is therefore very large. At large voltages the depletion of bound ions from one of the capacitor electrodes triggers a collapse of the capacitance to the mean-field value.

\end{abstract} \maketitle

The rising demand for compact forms of energy storage with high
power output has resulted in increased interest in
electrochemical capacitors (ECs) \cite{Abruna, Conway}. An EC is a
pair of metal electrodes separated by an ionic conductor, such as
an aqueous solution of ions, an ionic liquid \cite{Galinski}, a
super-ionic crystal \cite{Agrawal}, or an ion-conducting glass
\cite{Mariappan}.  ECs with extremely high area per unit
volume (``supercapacitors") are already used for a number of technologies.  In this paper our focus is not on large surface area, but on the deeper physical question of a maximum possible capacitance per unit area.

In a conventional double-plate capacitor, where metal electrodes
of area $S$ are separated by an insulator of width $d$ and
dielectric constant $\epsilon$, the capacitance $C = \epsilon S/4 \pi
d$ (in Gaussian units).  In an EC, the intervening medium is
actually a conductor with finite conductivity $\sigma$, but with
blocking of both ionic and electronic current at the electrode
interface.  The relation $C = \epsilon S/ 4 \pi d$ is therefore
valid only at sufficiently high frequencies $\omega \gg 4 \pi
\sigma / \epsilon$, when the bulk of the ionic medium behaves as an
insulator.  We concern ourself with the opposite limit $\omega \ll 4 \pi \sigma / \epsilon$, where polarization of the ionic medium eliminates electric field in the bulk and the capacitance of the EC is determined by the formation of thin electrostatic double layers (EDLs) at both electrodes.

How large can the capacitance be for these double layers?  The commonly-accepted expression for the maximum possible capacitance of an EDL goes back to Helmholtz \cite{Helmholtz1853}, who assumed that the charge of the metal surface is compensated by a layer of counterions with diameter $a$ residing on the surface of metal. The resulting ``Helmholtz
capacitance" is given by
\be 
C_H = \epsilon S/2 \pi a. \label{eq:Helmholtz} 
\ee
For a double-plate capacitor, where the EDLs formed at both
electrodes can be thought of as two equal capacitances connected in
series, the maximum capacitance is $C_H/2 = \epsilon S/4 \pi a$. For
$a = 2\textrm{\AA}$ and $\epsilon = 5$, as we use below, $C_H/2S = 22$ $\mu$F/cm$^2$.

A recent experiment \cite{Mariappan}, however, has reported much larger values of the EDL capacitance in phosophosilicate glasses placed between platinum electrodes.  Capacitance per unit area as large as $400\hspace{1mm} \mu \textrm{F/cm}^2$  was measured, corresponding to an effective capacitor thickness $d^*=\epsilon S/ 4 \pi C$ in the range $0.2$ -- $0.7$ \AA, much smaller than any ion radius. The glass was held at a temperature of 573 K, at which only the smallest ions, Na$^+$ with diameter $a = 2\textrm{\AA}$, are mobile.  The dielectric constant of the glass $\epsilon \approx 5$.

Current theories of EDL capacitors, based on the mean-field approach, fail to explain such large capacitance.  The most widely-used theory is that of Gouy, Chapman, and Stern (GCS) \cite{GCS}, which extends the Helmholz capacitor concept to allow for the thermal motion of counterions.  In this approach, neutralizing ionic charge is imagined as a stack of thin uniform layers placed parallel to the charged electrode, with the charge density of each layer dictated by the Poisson-Boltzmann equation. Such theories naturally lead only to a larger effective capacitor thickness, and therefore a smaller capacitance than the Helmholtz value.

In this paper we propose an alternate theory to explain the large
differential capacitance of the EDL. We abandon the mean-field
approximation and deal instead with discrete ion charges, which
interact strongly with the metal surface in a way that is not
captured by the mean-field approximation.  For simplicity, this paper will focus on the case where the conductivity $\sigma$ is due to a single species of ions with charge $e$.  Our conclusions may be applied to capacitors made with ion-conducting glasses or superionic crystals, where only the smallest positive ionic species (such as Na$^+$ or Li$^+$) is mobile. 

\begin{figure}[htb]
\centering
\includegraphics[width=0.45 \textwidth]{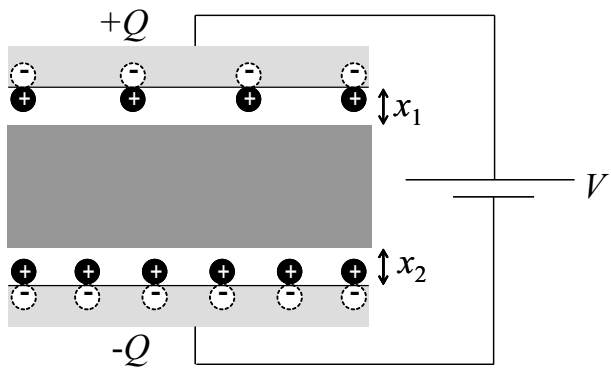}
\caption{A capacitor consisting of parallel metal plates (lightly shaded)
bounding an ionic conductor with mobile positive ions.  The neutral region of the ionic conductor is heavily shaded, while negative depletion regions ($x_1$ and $x_2$) are left white.} \label{fig:schematic}
\end{figure}

An ion adjacent to a metal electrode produces electronic polarization of the metal surface, and it experiences an attraction to the resulting image charge.  For ions of small radius, the image attraction is significantly larger than the thermal energy $k_BT$, so that ions form stable, compact ion-image dipoles at the metal surface.  Repulsion between adjacent dipoles results in the formation of a strongly-correlated liquid of dipoles along the surface of both electrodes (see Fig$.$ \ref{fig:schematic}).  When a voltage $V$ is applied between the electrodes, the build-up of an electronic charge $\pm Q$ on the electrodes drives ions to detach from the positive electrode and to bind to the negative one. Below, we demonstrate that the
resulting EDL capacitance $dQ/dV$ can be significantly larger than the
Helmholtz value $C_H/2$, since the dipole-dipole repulsion
that resists ion transfer is relatively small.

Consequently, our answer to the question in the title is that the
capacitance per unit area can be much larger than the Helmholtz value
$C_H/2S$.  In other words, the effective  effective capacitor
thickness $d^*$ can be much smaller than the ion radius. Below we derive an expression for $d^*$ that is reasonably close to experimental values. Note that we do not assume any Faradaic effects, as are employed in theories of so-called ``pseudo-capacitance" \cite{Conway}, to explain large apparent capacitance values.

Our theory also explains another peculiarity of the experiment
\cite{Mariappan}, namely the sharp drop of differential capacitance at a
certain critical voltage. We show that, indeed, capacitance should
achieve a maximum at a particular nonzero voltage before
collapsing to a much smaller value. Contrary to the mean-field
theory of Ref$.$ \cite{Kornyshev2007}, this maximum is not driven by
excluded volume effects among bound ions. Rather, the maximum occurs far below the complete filling of an ionic layer at either electrode \cite{Baldelli2008}, when the voltage difference induces the positively-charged electrode to lose all of its bound ions.

Here we consider the case of a parallel-plate capacitor, where the intervening medium is modeled as a fixed
negative background with charge density $-e N$, upon which resides
a neutralizing concentration of mobile positive ions with charge
$+e$ and bulk density $N$.  If we imagine the conducting ions to be hard spheres with diameter $a$, then an ion up against the metal surface experiences an attractive potential energy of approximately 
\be 
u_{im} \simeq -e^2/2 \epsilon a \label{eq:Uimage}. 
\ee 
For $T = 573$ K, $\epsilon = 5$, and $a = 2$ \r{A}, we get $|u_{im}|/k_BT \approx 15$, so that such ions are bound strongly to the surface. At a given voltage $V$, some area densities $n_1$ and $n_2$ of ions bind to the positive and negative plates, respectively.  The following arguments generally assume that $N a^3 \ll 1$.

The attachment of positive ions to the metal plates is associated with the formation of regions with negative net charge $q_1$ and $q_2$ near the anode and cathode, respectively. Each of these charges exactly cancels the net charge of the adjoining plate and its bound ions, so that there is no electric field in the bulk of the ionic conductor.  This implies $q_1 = -en_{1}S - Q$ and $q_2 = - en_{2}S +Q$, where $Q$ is the electronic charge that moves through the voltage source. In other words, the dipoles at each metal-glass interface are effectively embraced by a capacitor composed of the charge $q$ and its positive image $-q$ in the metal. As in every plane capacitor, the charges $q_1$ and $q_2$ are uniformly distributed along the plane.

Since positive ions gain a large energy $|u_{im}|$ by adsorbing to the metal surface, at equilibrium there must be a correspondingly large potential difference $V_{im} \equiv |u_{im}|/e$ between the metal surface and the bulk of the ionic conductor, so that the chemical potential of ions is uniform.  With such a large internal potential difference $V_{im} \gg k_BT/e$ at each electrode, the negative regions are strongly depleted of ions even at very small applied voltage $V$.  The charges $q_1$ and $q_2$ therefore constitute depletion layers of width $x_1$ and $x_2$ which form at the anode and cathode, respectively; here it is assumed that $x_1, x_2 \gg a$.  These layers are devoid of mobile ions and have a charge density equal to that of the negative background, so that $q_1 = -eNx_1$ and $q_2 = -eNx_2$.  Thus
\begin{eqnarray}
e N x_1 - e n_1 = Q/S, \label{eq:rightneutral} \\
e N x_2 - e n_2 = - Q/S. \label{eq:leftneutral}
\end{eqnarray}
The electrostatic energy associated with the formation of the depletion layers can be estimated as
\be
U_{dep} = \frac{2 \pi e^2 S} {3 \epsilon N} \left[ (n_1 + Q/Se)^3 + (n_2 - Q/Se)^3 \right] . \label{eq:Udep}
\ee

In addition to $U_{dep}$, there is a positive energy associated
with repulsion between bound ions on the metal surface.  When their
density is low enough that $n a^2 \ll 1$, ions repel
each other by a dipole-dipole interaction: the potential created
by a bound ion and its image charge repels an adjacent ion.  In
this limit the repulsive interaction between two adjacent dipoles is approximately
\be 
u_{dd}(n) = e^2 a^2 n^{3/2} / 2 \epsilon \label{eq:dipole}. 
\ee 
This repulsion results in the formation of a strongly-correlated liquid of ions on the electrode surface, reminiscent of a two-dimensional Wigner crystal.  The total dipole energy at a given plate is $\alpha n S u_{dd}(n)$, where $\alpha$ is a numerical coefficient which depends on the structure of the lattice of dipole positions.  For a triangular lattice, $\alpha \approx 4.4$.  Further calculations will use this value.

We now assemble a full description of the total energy $U$ associated with the bound charge densities $n_1$ and $n_2$.  Taking as our reference the case where $n_1 = n_2 = 0$, 
\be 
U = S(n_1 + n_2) u_{im} + U_{dep} + \alpha S \left[n_1 u_{dd}(n_1) + n_2 u_{dd}(n_2) \right] - Q V. \label{eq:Utot} 
\ee 
Here, $-Q V$ represents the work done by the voltage source. The equilibrium values of $Q(V)$, $n_1(V)$, and $n_2(V)$ are those which minimize $U$.  The condition $\partial U/\partial Q = 0$ gives
\be 
\frac{4 \pi}{N a^3} (n_1 a^2 + n_2 a^2)\left[\frac{2 Qa^2}{eS} - (n_2 - n_1)a^2 \right] = \frac{V}{V_{im}}. \label{eq:dUdQ}
\ee
Eq$.$ (\ref{eq:dUdQ}) suggests that at sufficiently small voltages, $Q \simeq eS(n_2 - n_1)/2$. Putting this result for $Q$ back into Eq$.$ (\ref{eq:Utot}) we get a total energy $U(n_1, n_2)$.  The conditions $\partial U/\partial n_1 = 0$ and $\partial U/\partial n_2 = 0$ produce (via their addition and subtraction) the following relations: 
\be 
\frac{\pi}{N a^3} (n_1 a^2 + n_2 a^2)^2 + \frac{5 \alpha}{4} \left[ (n_1 a^2)^{3/2} + (n_2 a^2)^{3/2} \right] = 1, \label{eq:nadd} 
\ee 
\be 
(n_2 a^2)^{3/2} - (n_1 a^2)^{3/2} = \frac{2 V}{5 \alpha V_{im}}.
\label{eq:nsub} 
\ee 

One immediate consequence of Eqs$.$ (\ref{eq:nadd}) and
(\ref{eq:nsub}) is that at zero voltage, when the net charge $Q$
of the capacitor is zero, there is still a finite concentration
$n_0$ of ions bound to each plate.  If ions are relatively sparse
in the bulk, so that $N a^3 \ll 1$, then the second term of Eq$.$
(\ref{eq:nadd}) can be neglected and 
\be 
n_0 a^2 \simeq \sqrt{N a^3/4 \pi } \ll 1.  \label{eq:n0}
\ee 
Eq$.$ (\ref{eq:n0}) implies that bound ions are sufficiently distant that a dipole-dipole interaction between them is justified.  It also ensures that the depletion layer widths $x_1, x_2 \gg a$, as assumed earlier.

As the voltage is increased from zero, ions are driven away from
the anode and attracted to the cathode, so that
$n_1$ decreases and $n_2$ increases.  By comparing Eqs$.$ (\ref{eq:Udep}) and (\ref{eq:dipole}), we see that the condition $N a^3 \ll 1$ implies that the energy cost associated with increasing the total number of bound ions, and thereby causing the depletion layers to swell, is much larger
than the dipole-dipole interaction energy.  As a consequence, the total
number of bound ions $n_1 + n_2 \simeq 2 n_0$ remains almost constant with increasing voltage. The electronic charge $Q \simeq e S (n_2 - n_1)/2$ that passes through the voltage source can therefore be thought of as the corresponding movement of image charges from one plate to
another. At a particular voltage $V_c$, however, the anode becomes completely depleted of bound ions.  This occurs when
$n_1 = 0$ and $n_2 \simeq 2 n_0$, so that Eq$.$ (\ref{eq:nsub}) yields
\be 
V_c \simeq \frac{5 \alpha}{2} \left( \frac{N a^3}{\pi} \right)^{3/4} V_{im}. \label{eq:Vc} 
\ee
Substitution of $n_1 + n_2 = 2 n_0$ into Eq$.$ (\ref{eq:dUdQ}) verifies that $Q \simeq e S(n_2 - n_1)/2$ for all $|V| < V_c$.
At $|V| > V_c$, the number of bound ions on the non-depleted electrode
may still increase, but only through the costly widening of the
depletion layer.

Knowing $n_1$ and $n_2$ as a function of voltage allows us to
calculate the differential capacitance 
\be 
C(V) = \frac{dQ}{dV}
\simeq \frac{e S}{2} \left( \frac{dn_2}{dV} - \frac{dn_1}{dV} \right). 
\label{eq:Cdef}
\ee
Taking the derivative of Eq$.$ (\ref{eq:nsub}) with respect to $V$, and using $n_1 + n_2 = 2n_0$, gives
\be 
C(0) \simeq
\frac{4}{15 \alpha} \left( \frac{4 \pi }{N a^3} \right)^{1/4} \frac{\epsilon S}{a} = \frac{8 \pi}{15 \alpha} \left( \frac{4 \pi}{N a^3} \right)^{1/4} C_H \label{eq:C0}
\ee
for the capacitance at zero voltage. For $Na^3 \ll 1$ we arrive at $C(0) \gg C_H/2$.  Such large capacitance arises because at $V < V_c$ charging of the capacitor is limited only by the dipole-dipole repulsion energy, which is small since $n_0 a^2 \ll 1$.

Note that at $V = 0$ the capacitor is composed of two identical EDLs in series, so that the total capacitance $C(0)$ is equal to half the capacitance of each. At higher voltages, the capacitance of the
EDL near the positive plate increases strongly as this plate becomes depleted of ions and the corresponding dipole-dipole energy goes to zero.  Thus the contribution of the positive plate to the total capacitance vanishes at $V = V_c$.  As a result, immediately prior to $V = V_c$ the capacitance achieves its maximum value 
\be 
C_{max} \simeq \frac{8}{15 \alpha} \left( \frac{\pi}{N a^3 } \right)^{1/4} \frac{\epsilon S}{a} = \sqrt2 C(0) . 
\ee 
At $V > V_c$ the capacitance collapses to a much
smaller value.  This ``depleted capacitance" can be
found through optimization of the total energy $U$ under the
condition $n_1 = 0$, which yields 
\be 
C_{dep}(V) \simeq  \frac{\epsilon S}{a} \sqrt{\frac{N a^3 }{4 \pi (V/V_{im} + 1)} } . \label{eq:Cdep}
\ee
This expression neglects the weak dipole-dipole interaction at the non-depleted negative plate.  At $V/V_{im} \gg 1$, the capacitance is dominated by the depletion layer next to the positive electrode, and therefore it approaches the standard value for capacitance of a depletion layer.

\begin{figure}[htb]
\centering
\includegraphics[width=0.45 \textwidth]{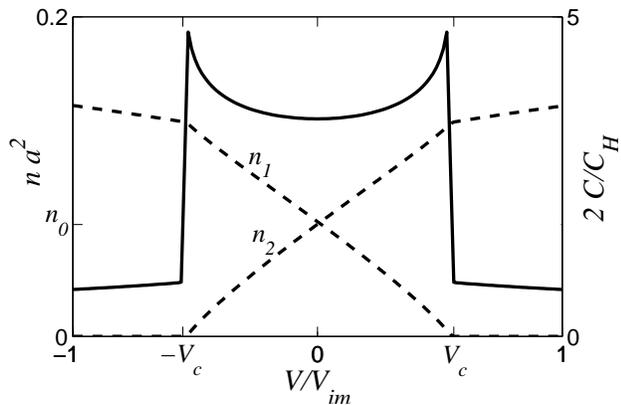}
\caption{The area densities of bound ions (dashed lines, left vertical axis) and the capacitance (solid line, right vertical axis) as a function of applied voltage for $Na^3 = 0.1$.} \label{fig:c-n}
\end{figure}

Fig$.$ \ref{fig:c-n} shows the capacitance and the density of bound ions as a function of voltage, as calculated by a numerical minimization of the total energy in Eq$.$ (\ref{eq:Utot}).  We have used $Na^3 = 0.1$, following the estimate of Ref$.$ \cite{Mariappan}.

The effective thickness $d^*_{min}$ corresponding to $C_{max}$ is
\be 
d^*_{min} \simeq  \frac{15 \alpha}{32 \pi} \left( \frac{N a^3}{\pi} \right)^{1/4} a \approx 0.49 \left(Na^3\right)^{1/4} a . 
\ee 
Thus we arrive at a remarkable prediction: the effective capacitor thickness can be much smaller than the ion diameter $a$. As an example, an ion-conducting glass with mobile sodium atoms of diameter $a = 2$
\r{A} and density $N a^3 = 0.1$ can be used to make a capacitor with capacitance nearly five times larger than $C_H/2$. As mentioned before, this surprisingly high capacitance is a result of the weak dipole-dipole interaction between bound ions that comprise the double layer.  Indeed, near the capacitance maximum the filling factor on the negative plate $n_2 a^2 \simeq 2 n_0 a^2 \ll 1$, so that it is incorrect to think of the EDL as a series of uniformly charged layers. This difference represents an important change of paradigm, from a mean-field capacitor to a capacitor composed of discrete, correlated dipoles.

\begin{figure}[htb]
\centering
\includegraphics[width=0.45 \textwidth]{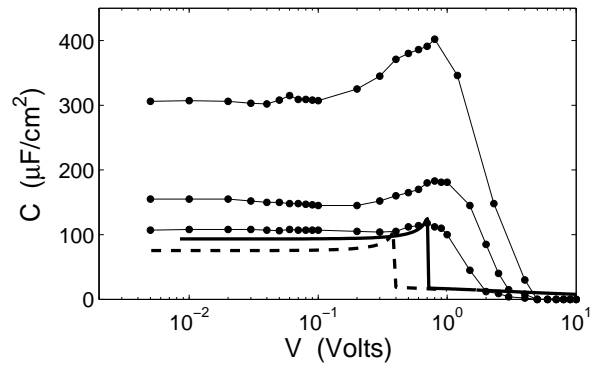}
\caption{The capacitance of an ion-conducting glass between metal
plates.  The dashed and solid lines are theoretical predictions, calculated by a numerical minimization of the energy in Eq$.$ (\ref{eq:Utot}).  The dashed line uses $\epsilon = 5$ while the solid line assumes $\epsilon = 10$, $\epsilon_m = 2$.  Thin lines with dots show data for three different glasses from Ref$.$ \cite{Mariappan}.}
\label{fig:compare_to_glass}
\end{figure}

We can compare our theory to the experiments of Ref$.$ \cite{Mariappan}. Capacitance-voltage characteristics for three different phosphosilicate glasses are shown in Fig$.$ \ref{fig:compare_to_glass}
together our theoretical prediction (dashed line), using $\epsilon = 5$ and $N = 10^{22}$ cm$^{-3}$ as estimated by the authors of Ref$.$ \cite{Mariappan}. The agreement with theory can be improved (solid line) if we assume that the image attraction of Eq$.$ (\ref{eq:Uimage}) uses a smaller dielectric constant $\epsilon_m = 2$ \cite{kappanote}, while the macroscopic dielectric constant $\epsilon = 10$.  

While our theory seems to provide a qualitative explanation of the data, the observed collapse is not as sharp as we predict.  This may be due in part to the fact that our theory is valid only in the limit $Na^3 \ll 1$, and the use of a marginally small $Na^3 = 0.1$ does not allow for a large separation between the length scales $a$, $n_0^{-1/2}$, and $x_1$, $x_2$.  We have also ignored temperature and disorder effects.  The importance of thermal effects can be estimated by calculating the mean-square thermal displacement $\langle r^{2} \rangle^{1/2} \equiv \delta$ of a given dipole from its potential energy minimum.  For $\epsilon = 5$ and $T = 573$ K, we get relatively small displacements $\delta \simeq 0.3n^{-1/2}$ at $n = n_0$ and $\delta \simeq 0.2n^{-1/2}$ at $n = 2n_0$.  Thus it seems unlikely that thermal motion alone can be responsible for the substantial smearing of the capacitance collapse.

One can imagine that at much higher temperatures, such that $k_BT \gg u_{dd}(n_0)$ but $k_BT \ll u_{im}$, the thermal energy of bound ions becomes much larger than their mutual repulsion, and therefore dipoles on a given metal surface are better described as a two-dimensional ideal gas than as a Wigner crystal.  In this limit the capacitance is determined by the change in the ideal gas entropy associated with transfer of ions from one electrode to the other.  This description gives a zero-voltage capacitance $C(0) = n_0 a^2 (u_{im}/k_BT) (\epsilon S/a)$. For the crossover temperature $T_c$ to Eq$.$ (\ref{eq:C0}) we get $T_c = 15 \alpha u_{dd}(n_0)/4 k_B \approx 3700$ K.  Therefore, at $T = 573 \textrm{ K}$, deviations from our low-temperature theory are unlikely to be large.  The disorder potential acting on mobile ions in the glass may be more important, and we leave this for the subject of a later publication.

We are grateful to A. L. Efros, M. M. Fogler, J. Schmalian, T. T. Nguyen, S. D. Baranovskii, and B. Roling for helpful discussions.  B. S. acknowledges the support of the NSF and M. S. L. thanks the FTPI for financial support.


\end{document}